\documentclass{raa}
\usepackage{graphicx,times}
\usepackage{natbib}
\usepackage{amssymb,amsmath}
\usepackage[T1]{fontenc}
\usepackage{aecompl}
\usepackage{threeparttable}
\bibpunct{(}{)}{;}{a}{}{,}
\usepackage[a4paper=true,dvipdfm=true,pagebackref=true]{hyperref}
\hypersetup{pdftitle = GeneticKNN: A Weighted KNN Approach Supported by Genetic Algorithm for Photometric Redshift Estimation of Quasars, pdfauthor = Han B. et al., pdfsubject= The subject, pdfkeywords = keyword1 keyword2 keyword3}
\hypersetup{colorlinks = true, linkcolor = green, anchorcolor = red, citecolor = blue, filecolor = red, pagecolor = red, urlcolor = red}

\begin{document}
\title{GeneticKNN: A Weighted KNN Approach Supported by Genetic Algorithm for Photometric Redshift Estimation of Quasars$^*$ \footnotetext{\small $*$ Supported by the National Natural
Science Foundation of China.} }

 \volnopage{ {\bf 2020} Vol.\ {\bf X} No. {\bf XX}, 000--000}
   \setcounter{page}{1}

   \author{Bo Han\inst{1}, Li-Na Qiao\inst{1}, Jing-Lin Chen\inst{1}, Xian-Da Zhang\inst{2}, Yanxia Zhang\inst{3}, Yongheng Zhao\inst{3}}
   \institute{School of Computer Science, Wuhan University, Wuhan 430072, China\\
   Institute of Electronics, Chinese Academy of Sciences, 100190, Beijing, China \\
    CAS Key Laboratory of Optical Astronomy, National Astronomical
Observatories, Beijing, 100101, China {\it zyx@bao.ac.cn}\\
\vs \no
   {\small Received 2019 July; accepted 2020 }
}

\abstract{We combine $K$-Nearest Neighbors (KNN) with genetic algorithm (GA) for photometric redshift estimation of quasars, short for GeneticKNN, which is a weighted KNN approach supported by GA. This approach has two improvements compared to KNN: one is the feature weighted by GA; another is that the predicted redshift is not the redshift average of $K$ neighbors but the weighted average of median and mean of redshifts for $K$ neighbors, i.e. $p\times z_{\rm median} + (1-p)\times z_{\rm mean}$. Based on the SDSS and SDSS-WISE quasar samples, we explore the performance of GeneticKNN for photometric redshift estimation, comparing with the other six traditional machine learning methods, i.e. Least absolute shrinkage and selection operator (LASSO), support vector regression (SVR), Multi Layer Perceptrons (MLP), XGBoost, KNN and random forest. KNN and random forest show their superiority. Considering the easy implementation of KNN, we make improvement on KNN as GeneticKNN and apply GeneticKNN on photometric redshift estimation of quasars. Finally the performance of GeneticKNN is better than that of LASSO, SVR, MLP, XGBoost, KNN and random forest for all cases. Moreover the accuracy is better with the additional WISE magnitudes for the same method.
}

   \authorrunning{B. Han et al. }            
   \titlerunning{GeneticKNN}  
   \maketitle

\section{INTRODUCTION}
In the era of full-band astronomy, with the huge amount of photometric data or images available, photometric redshits play more important role than ever to measure the distances of the celestial bodies, which are of great significance to the study of many fields of astronomy. Compared with traditional spectral redshifts, photometric redshifts have the advantages of high efficiency, low cost and can be applied to a large number of celestial objects with minimal telescope observation time. Especially, they provide unique values for those faint sources whose spectral information can been obtained.

At present, many methods have been successfully applied to estimate photometric redshifts. These approaches can be classified into two categories: template fitting method and machine learning models.

Template fitting method is a traditional method to estimate redshifts by extracting features, such as multiband values, from celestial observations and matches them to design templates constructed from theoretical models or real sample observations. This type of methods have the advantages of small calculation cost and easy implementation. Bolzonella et~al. (2000) used standard spectral energy distribution (SED) of wavebands to estimate redshifts. Wu et~al. (2004) applied the empirical color-redshift relation to minimize $\chi^{2}$ function for estimating redshifts of quasars. Rowan-Robinson et~al. (2008) applied fixed galaxy and quasar templates to the photometric data of 0.36-4.5$\mu$m and used a set of four infrared emission templates for infrared excess data of 3.6-170$\mu$m. IIbert et~al. (2009) applied SED (Le Phare) to compute the redshift in 2-deg$^{2}$ COSMOS. The series of the above research results suggested that the estimation accuracy of SED approaches highly depends on the mapping to the designed templates. These templates are constrained by direct mapping to the theory simulation results or sample observation data.

Machine learning models apply statistics and machine learning algorithms on training data sets to automatically learn the complex functional correlation between multi-band photometric observations and their corresponding redshifts. These algorithms are data-driven, different from template-driven. Ball et~al. (2007) used an nearest neighbor algorithm to estimate the redshifts of galaxies and quasars. Abdalla et~al. (2008) proposed a neural network approach. Freeman et~al. (2009) came up with a nonlinear spectral connectivity analysis method to convert photometry color to a simpler and more natural coordinate system, and used regression to estimate photometric redshifts. Gerdes et~al. (2010) developed an ArborZ enhanced decision tree algorithm for photometric redshift estimation. Way et~al. (2012) proposed a method based on self-organizing mapping (SOM). Bovy et~al. (2012) developed the extreme deconvolution technique and confirmed that the increase of ultraviolet and near-infrared band information was very important for photometric quasar-star separation and redshift estimation. Mortlock et~al. (2012) developed Bayesian model comparison techniques to perform probabilistic selection of high-redshift quasars based on SDSS and UKIDSS databases. Carrasco et~al. (2013) designed an algorithm of using prediction tree and random forest algorithm. It tolerates measurement errors in calculations and support effectively processing missing values of photometric data. Brescia et al. (2013) applied a neural network algorithm (MLPQNA) to estimate photometric redshifts of quasars based on datasets from four different projects (SDSS, GALEX, UKIDSS and WISE). Schindler et~al. (2017) introduced a random forest algorithm to solve the uncertainty of photometric redshift estimation.
In order to further improve the redshift estimation accuracy, researchers considered some novel methods or the combination of several methods. Wolf (2009) integrated $\chi^{2}$ template fitting and empirical training in a framework and applied them to improve the qusar estimation accuracy on SDSS DR5 dataset. Laurino et~al. (2011) proposed a Weak Gating Expert (WGE) approach through combination of multiple data mining techniques to measure the redshift of galaxies and quasars. Gorecki et~al. (2014) studied different methods and combined SED and machine learning approaches for galaxy redshift estimation. Han et~al. (2016) integrated $K$-nearest neighbor (KNN) algorithm and Support Vector Machine (SVM) for correcting quasar redshift estimation biases. Hoyle (2016) proposed a deep neural network to estimate galaxy redshifts by using full galaxy images in each measured band. Leistedt \& Hogg (2017) presented a new method for inferring redshifts in deep galaxy and quasar measurements which constructed template SED directly from spectral training data and combined the advantages of machine learning method and template fitting method. Wolf et~al. (2017) studied the redshift properties of several experiential and template methods and drew a conclusion that Kernel Density Estimation (KDE) achieved the best results. Jouvel et~al.(2017) compared SED, a neural network and a random forest methods, and  explored different technologies to reduce the redshift estimation outlier fraction. Speagle \& Eisenstein (2017a, b) computed photometric redshifts by using fuzzy prototyping and self-organizing mapping methods. They showed that estimation robustness and flexibility could be obtained by combining template fitting and machine learning methods. Their exploration provided useful insights for astronomers to further explore of the color-redshift relationship. Dsanto \& Polsterer (2018) explored deep Learning methods to derive the redshifts with confidence probabilistic metering directly from the multi-band imaging data.

In recent research, the most accurate redshift estimation was obtained by $K$-nearest neighbors (KNN) and Random Forest (RF) approaches (Zhang et~al. 2019). Between them, RF algorithm has better results on the SDSS dataset, while its estimation on the SDSS-WISE dataset is not as good as KNN algorithm. Meanwhile, RF algorithm, as an integrated algorithm, is difficult to optimize and takes a long time to train. Especially when we choose mean absolute error (MAE) as the optimization index, the training time shows a nonlinear growth. On the contrary, KNN algorithm is simple and achieves high estimation accuracy. However, KNN algorithm still has some drawbacks. For example, Euclidean distance in the color space is usually computed to infer nearest neighbors. But the distance calculation does not consider different level of attribute contributions from different bands and assign them equal weights. Nevertheless the performance of KNN may be improved by integration with other methods. For example, KNN was designed with genetic algorithm (GA) or decision tree (DT), or combined with GA and DT, and applied in different fields (Houben et~al. 1997 and references; Suguna \& Thanushkodi 2010; Yan et~al. 2013; Mclnerney et~al. 2018).

For solving the above problem, we firstly apply a novel attribute weighted KNN method to improve the estimation accuracy for photometric redshifts of quasars. Specifically, we use genetic algorithm to search for the optimal weights, and then perform KNN for the computation of quasar photometric redshifts. Based on the data distribution and location analysis, it assigns the weighted average of the mean and the median of redshifts as the predicted redshifts for KNN algorithm. The overall structure of the study takes the form of five sections. Section~2 describes the datasets used in our study, and Section~3 gives the detailed algorithm of GeneticKNN. The experimental results are reported in Section~4. Finally, Section~5 draws a conclusion of our study.

\section{Samples}
Sloan Digital Sky Survey (SDSS; York et~al. 2000) obtains huge amounts of photometric and spectral information for celestial bodies in the Universe. The quasar sample we adopt is from the data release 14 Quasar catalog (DR14Q) of the SDSS-IV/eBOSS (Paris et~al. 2018). The DR14Q catalog contains 526,356 unique quasars, of which 144,046 are new discoveries since the beginning of SDSS-IV. And it also includes previously spectrally confirmed quasars from SDSS-I, II and III. For SDSS-I/II/III, spectral observations of quasars were conducted on 9,376~deg$^{2}$ which for the new SDSS-IV, spectral observations of quasars were above 2,044~deg$^{2}$ The number of the SDSS quasar sample reduces to 445,958 after removing the records which contain default SDSS magnitudes, $\emph{zWarning} = -1$ and full magnitude errors large than 5.In our study, we used $u_{\rm AB}$ = $u'- 0.04mag$ and $z_{\rm AB} = z' + 0.02mag$ to covert the SDSS $u-band$ and $z-band$ magnitudes to AB magnitudes. All magnitudes for galactic extinction can be corrected by the use of the extinction values from the DR14Q catalog. We directly acquire W1 and W2 from the DR14Q catalog which provides the information from SDSS and WISE (Wide-field Infrared Survey Explorer, Mainzer et~al. 2011) datasets. When the records with missing W1 and W2 are excluded, the number of the SDSS-WISE quasar sample reaches 324,333. W1 and W2 are converted to AB magnitudes by using $W1_{AB} = W1 + 2.699$ and $W2_{\rm AB}+3.339$ and extinction-corrected by the extinction coefficients $\alpha_{w1} = 0.189$ and $\alpha_{w2} = 0.146$ with the extinction values from the SDSS photometry. AB magnitude conversion and reddening correction process is just like the work of Schindler et al. (2017).

For SDSS quasar sample, we consider two input patterns: five magnitudes-four colors (5m\_4c) with $u,g,r,i,z,u-g,g-r,r-i,i-z$ and $r$ magnitude-colors (r\_4c) with $r,u-g,g-r,r-i,i-z$. And for SDSS-WISE quasar sample, we also think about two input patterns: 7 magnitudes-6 colors (7m\_6c) with $u,g,r,i,z,W1,W2,u-g,g-r,r-i,i-z,z-W1,W-W2$ and $r$ magnitude-6 colors (r\_6c) with $r,u-g,g-r,r-i,i-z,z-W1,W1-W2$.

\section{The Principle of GeneticKNN Algorithm}
KNN method is a regression estimation algorithm based on distance, which is quite sensitive to distance. However, it still has many problems not considered, such as using equal weights for features rather than considering feature weights and the relationship between the density distribution of nearest neighbors and locations. We consider a new method based on KNN weighting the data to improve the accuracy of redshift estimation. This attribute weighting method can not only solve the problem of inconsistent value range of data, but also adjust the weights of different attributes to further solve the problem of collinearity of attributes. At the same time it can assign large weights to attributes that play an important role in photometric redshift estimation. Therefore, the choice of weight is a very complicated problem.

Genetic algorithm is a heuristic algorithm proposed by Holland (1992) inspired by the process of natural selection, and it is a kind of evolutionary algorithm. Genetic algorithms rely on biologically inspired operational methods such as mutation, crossover, and selection, which are often used to generate high-quality solutions to optimization and search problems. Although it may not be able to find the optimal solution, it can find a better solution. Therefore, the genetic algorithm may not be in an optimal position at some time, but it can make the gene change and jump to a better position through mutation. A key feature of genetic algorithm is that it keeps the entire population evolving, and even if an individual loses a useful trait, the rest of the population will retain that useful trait and continue to improve. Since genetic algorithm only needs to master the information of the target function and does not need to understand the requirements of continuous differentiability, derivative and other functions, it has good adaptability and can be applied to solve various problems (Man et~al. 2012). Genetic algorithm mainly solves optimization problems. If modeled properly, genetic algorithm can solve most of the optimization problems that arise in practice. In this chapter, we explore KNN algorithm based on heredity, and search the optimal weights of features by genetic algorithm, so as to improve the accuracy of redshift estimation.

The study of machine learning combined with genetic algorithm is a new research goal, which generalizes genetic algorithm from traditional discrete space to a new machine learning algorithm with special generation rules. This learning idea improves the problem of knowledge condensation and knowledge extraction in artificial intelligence, and provides ideas for many problems that are in deadlock. Genetic algorithm, which is part of search algorithms, is all about machine learning.
The combination of genetic algorithm and other techniques is becoming more common. Some researchers proposed a method based on genetic algorithm to select a group of optimal features to build a classifier model, such as Decision Tree (DT), Naive Bayes (NB), K-Nearest Neighbor(KNN) and SVM, etc., and established GA-SVM Hybrid Model to obtain two optimal subsets. Silva multi-objective genetic algorithm constructs sparse Least Squares Support Vector Machine (LS-SVM) classifier.
It is a challenging task to find the best feature weight in the huge search space. We design a new two-step method based on genetic algorithm and K-Nearest Neighbor algorithm. Firstly, we use KNN algorithm to guide the genetic algorithm after many optimization to search for the optimal weight, and then the weight is combined with KNN model to complete the regression prediction. Unlike statistical methods, GeneticKNN model doesn't require information about feature weights; instead, it receives feedback from KNN models to determine the search direction. At the same time, we combine the predicted value of the algorithm with the research on the distribution density and location based on the nearest neighbor value, and put forward the overall lifting algorithm combining the attribute weighting and the weighted average based on the median and the average. The algorithm depends on the randomness of genetic algorithm and the ability of using fitness function to make the population converge to the expected point and combines a feedback concept similar to Neural Network to improve the estimation accuracy.
By weighting features, the algorithm not only adjusts the value range of data, but also reflects the importance of features and the correlation between features and real values. The algorithm flow chart is shown as Figure~1.

\begin{figure}
	\centering
	\includegraphics[width=\linewidth]{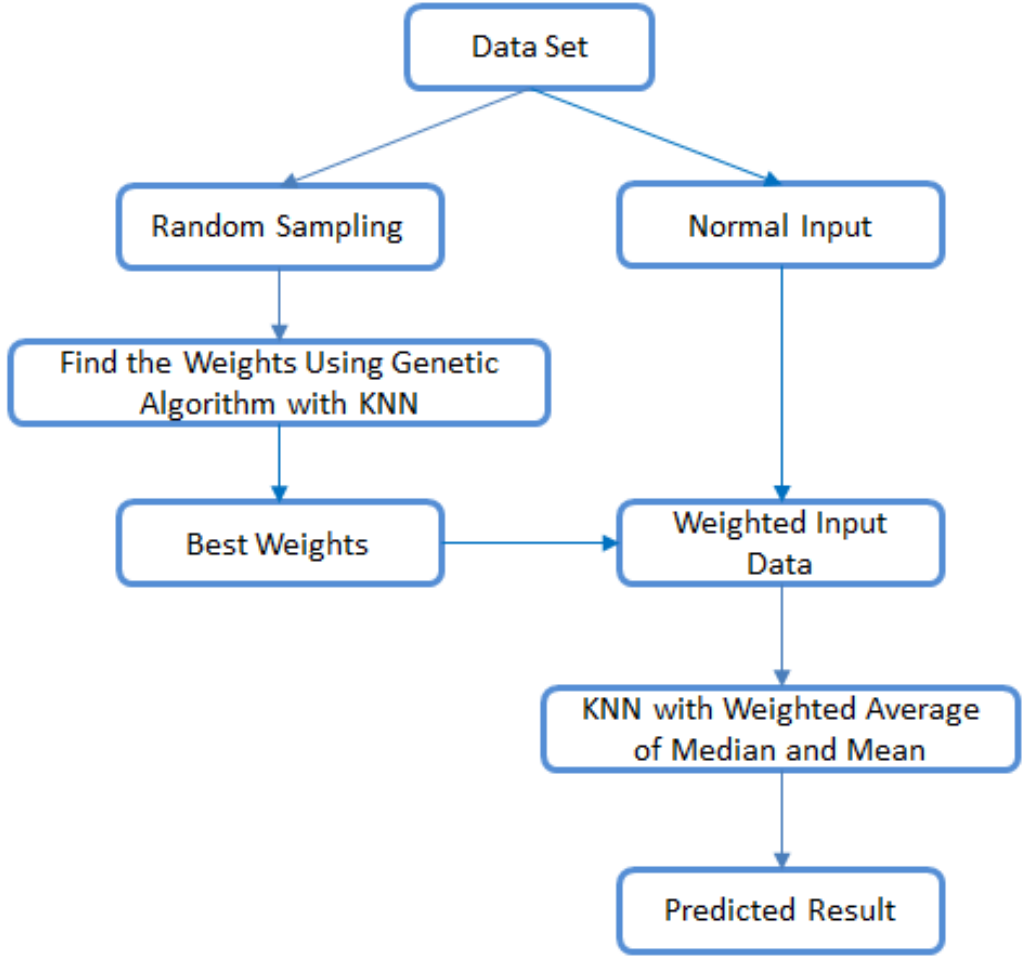}
	\caption{The algorithm flow chart of GeneticKNN.}
	\label{fig:GeneticKNN}
\end{figure}

The two most important designs of GeneticKNN algorithm are using genetic algorithm to search for feature weights and estimation optimization based on nearest neighbor value. In the Appendix, we introduce the design and implementation of the two important parts of GeneticKNN algorithm. The Evaluation Indexes for Photometric Redshift Estimation are detailed in in Appendix C.

\section{Results and Discussion}
In general, data processing is important, we explore how various preprocessing methods affect the accuracy of photometric redshift estimation. The experimental results show that normalization of data, feature selection (LASSO) and feature extraction (principal component analysis, PCA) are not helpful to redshift estimation for our cases. We don't present the detailed process here for saving space. All the following experiments were carried out in the Python 3.6 and scikit-learn 0.19.2 environment running on a workstation with an Inter(R) Core(TM) i5-4210U (1.70 GHz) CPU with 8 GB memory, Windows 10 system. XGBoost (version 1.1.0) is adapted from https://xgboost.ai/.

For the SDSS and SDSS-WISE quasar samples, we carry out photometric redshift estimation by linear model LASSO (Least absolute shrinkage and selection operator) algorithm, SVR (support vector regression), Multi Layer Perceptrons (MLP), XGBoost, KNN ($K$-Nearest Neighbors), Random Forest and GeneticKNN, evaluate all the models by 10-fold cross validation and select the models with better performance and less running time. Here MLP adopts a three layer structure: input layer, one hidden layer, output layer. The evaluation indexes are $\delta_{0.1}$, $\delta_{0.2}$, $\delta_{0.3}$ , $\sigma$ , MSE, $R^2$, Time. MAE is chosen as the optimization target of model training because four metrics are related to absolute error. However, for Random Forest, MSE as a standard is equivalent to reducing the variance as feature selection, using each terminal node to minimize the loss of L2 (Least Square Errors). MAE stands for mean absolute error and uses the median value of each terminal node to minimize the loss of L1 (Least Absolute Deviations). With the increase of samples, the running time of MSE does not change greatly, while that of MAE increases. The running time of MAE is $O(N^2)$. We construct a sorted list by adding elements consecutively to a sorted list and add elements by putting every element at the end and bubbles it to the right place on the list. It takes $O(N^2)$ to implement the sorting of the median. Updating the loss function is more difficult when the standard is MAE (and therefore the Random Forest output is median instead of average). For MSE, we adopt the original variance and perform a constant time update of the loss as we move the sample from one side of the partition to the other. While for MAE, the current loss needs to be completely recalculated, which spends $O(N^2)$. As a result, the whole process is $O(N^2)$ rather than $O(N)$. When the number of samples is small, Random Forest performs very well. But when the samples increase, the running time would increase nonlinearly. It takes a lot of time on training. Therefore, RF model is trained by MSE, and other models are trained by MAE. SVR algorithm is only applicable to small samples, so we sample the data and apply SVR algorithm. For comparison, the parameter settings of KNN algorithm and RF algorithm are consistent with those in Zhang et~al. (2019). The $K$ value of KNN algorithm is 30 in any experimental setting in this paper. The parameters of RF algorithm on the SDSS dataset are set as $n\_estimators=300$, $max\_depth=15$, while on the SDSS-WISE dataset, $n\_estimators=300$, $max\_depth=20$. We improve KNN as GeneticKNN from feature weighting and model modulation, feature weighting is by genetic method, the predicted redshift is the weighted average of median and mean of redshifts for $K$ neighbors, i.e. $p\times z_{\rm median} + (1-p)\times z_{\rm mean}$. For GeneticKNN, through KNN with feature weight by genetic algorithm, the optimal weights of $u,g,r,i,z,u-g,g-r,r-i,i-z$ are 0.8929, 1.5401, 0.5668, 1.9808, 0.9211, 9.3280, 9.4386, 9.6603, 6.4470, respectively; those of $r,u-g,g-r,r-i,i-z$ are -1.5862, 3.9205, 7.9372, -8.5756, -8.8144, respectively; those of $u,g,r,i,z,W1,W2,u-g,g-r,r-i,i-z,z-W1,W1-W2$ are 0.4240, 1.3062, 1.1441, -2.3406, 0.2213, 0.7625, 3.2117, 6.3703, 9.1243, -9.7937, -7.7430, -3.3514, 7.7508, respectively; those of $r,u-g,g-r,r-i,i-z,z-W1,W1-W2$ are 2.2635, 4.6948, -5.8334, 8.7845, -8.5422, 3.5702, 4.2004, respectively. After experiments, the parameter $p$ has the following settings: $p=0.3$ for the SDSS sample with 5m\_4c input pattern; $p=0.4$ for the SDSS sample with r\_4c input pattern; $p=0.5$ for the SDSS-WISE sample with 7m\_6c input pattern; $p=0.6$ for the SDSS-WISE sample with r\_6c input pattern.

For the SDSS quasar samples, r\_4c and 5m\_4c are taken as input patterns for all methods, the estimated results are shown in Tables~1-2, respectively. For the SDSS-WISE quasar samples, 7m\_6c and r\_6c
are taken as input patterns for all methods, the estimated results are indicated in Tables~3-4, respectively. First of all, from the standpoint of input pattern, we make performance comparison: for the SDSS dataset with 5m\_4c and r\_4c, it is seen in Tables~1-2 that in terms of $\delta_{0.1}$, the best performance with r\_4c is 0.1\% better than that with 5m\_4c, while in terms of $\delta_{0.2}$, $\delta_{0.3}$, $\sigma$, MSE and $R^2$, it is 0.01\% worse than that with 5m\_4c, but for running time, it takes less time; as for the SDSS-WISE dataset with 7m\_6c and r\_6c, it is found in Tables~3-4 that the best performance with 7m\_6c is slightly superior to that with r\_6c in all evaluation indexes, but it takes more time. In the meanwhile, the results with the input patterns for the SDSS-WISE sample are much better than that for the SDSS sample. Therefore, we could greatly improve the accuracy of prediction by increasing information from more bands. Then, from the algorithm perspective, the results reveal that the LASSO is a linear model which has the worst performance but the least running time. The results of SVR and MLP are the next. SVR shows poor performance on regression and has a relatively long running time. MLP model has many model parameters, takes a long time, and its accuracy is relatively low. XGBoost, RF and KNN have better performance, and the prediction accuracy of the three algorithms is not much different.
The detailed analysis of performance indicates that the linear regression model performs poorly, which is consistent with the fact that the redshift estimation is a nonlinear regression problem. Compared with other algorithms, the prediction accuracy of RF and XGBoost in the tree model is relatively high in some input patterns, but the running time is the longest. In addition, RF and XGBoost models are integrated models, and training and optimization of models takes a long time and are not easy to promote. KNN model comparatively has the advantages of high prediction accuracy and short running time, and can be further improved as GeneticKNN. For all cases, GeneticKNN is the best regressor for photometric redshift estimation. Compared with KNN, the performance of GeneticKNN has significant improvement both for the SDSS and SDSS-WISE datasets with any input pattern on $\delta_{0.1}$, $\delta_{0.2}$, $\delta_{0.3}$ and $\sigma$. The effect is most noticeable with 5m\_4c input pattern and it improves by 4.3\%, 1.8\%, 0.54\% and 1.75\% separately on indicators above. For r\_4c input pattern, it increases by 2.54\%, 1.09\%, 0.28\% and 1.01\%; for 7m\_6c, it promotes by 3.59\%, 1.35\%, 0.38\% and 1.69\%; for r\_6c, the improvements are 3.93\%, 1.48\%, 0.22\% and 1.58\%, respectively. Compared with RF, for 5m\_4c, the accuracy of GeneticKNN enhances by 3.19\%, 1.26\%, 0.11\% and 0.94\%; for r\_4c, it ascends by 1.87\%, 0.76\%, 0.05\% and 0.53\%; for 7m\_6c, it rises by 2.4\%, 0.92\%, 0.21\% and 0.94\%; and for r\_6c, 4.03\%, 1.55\%, 0.27\% and 1.56\%, respectively.

\begin{table*}
\begin{threeparttable}
\caption{Performance of photometric redshift estimation of different models for the SDSS sample with r\_4c}
\begin{tabular}{lccccccc}
\hline
Algorithm & $\delta_{0.1}(\%)$  & $\delta_{0.2}(\%)$ & $\delta_{0.3}(\%)$ & $\sigma$ & MSE & $R^2$ & Time(s) \\
\hline
LASSO     & 32.45  & 73.26  & 82.06  & 0.4976 & 0.3777 & -0.6991 & \textbf{3}       \\
SVR       & 62.03  & 80.45  & 84.65  & 0.3660 & 0.2913 & 0.0596  & 1214    \\
MLP        & 61.02  & 79.69  & 86.54  & 0.3417 & 0.2392 & 0.3487  & 3834    \\
XGBoost   & 62.07  & 80.20  & 87.32  & 0.3313 & 0.2286 & 0.3875  & 2172    \\
KNN       & 62.72  & 80.20  & 87.16  & 0.3311 & 0.2331 & 0.3681  & 95      \\
RF        & 63.39  & 80.53  & 87.39  & 0.3263 & \textbf{0.2283} & 0.3875  & 9584    \\
GK &\textbf{65.26}&\textbf{81.29}&\textbf{87.44}&\textbf{0.3210}&0.2325&\textbf{0.3948}&93\\
\hline
\end{tabular}
\begin{tablenotes}
\item[*] The definition of $\delta_{0.1}$, $\delta_{0.2}$, $\delta_{0.3}$, $\sigma$, MSE and $R^2$ refers to Appendix C.
\end{tablenotes}
\end{threeparttable}
\end{table*}

\begin{table*}
\caption{Performance of photometric redshift estimation of different models for the SDSS sample with 5m\_4c}
\begin{tabular}{lccccccc}
\hline
Algorithm & $\delta_{0.1}(\%)$  & $\delta_{0.2}(\%)$ & $\delta_{0.3}(\%)$ & $\sigma$ & MSE & $R^2$ & Time(s) \\
\hline
LASSO & 32.41 & 73.22 & 82.05 & 0.4977 & 0.3777 & -0.6983 & \textbf{115}     \\
SVR   & 60.81 & 79.76 & 84.33 & 0.3709 & 0.2933 & 0.0732  & 1403    \\
MLP    & 59.90 & 79.11 & 86.70 & 0.3475 & 0.2411 & 0.3286  & 3834    \\
XGBoost & 62.30 & 80.27 & 87.41 & 0.3303 & 0.2281 & 0.3908  & 2819    \\
KNN  & 62.18 & 80.00 & 86.99 & 0.3344 & 0.2353 & 0.3512  & 137     \\
RF    & 63.29 & 80.54 & 87.42 & 0.3263 & \textbf{0.2277} & 0.3887  & 16574   \\
GK &\textbf{66.48}&\textbf{81.80}&\textbf{87.53}&\textbf{0.3169}&0.2340&\textbf{0.4016}&\textbf{115}\\
\hline
\hline
\end{tabular}
\end{table*}

\begin{table*}
\caption{Performance of photometric redshift estimation of different models for the SDSS-WISE sample with r\_6c}
\begin{tabular}{lccccccc}
\hline
Algorithm & $\delta_{0.1}(\%)$  & $\delta_{0.2}(\%)$ & $\delta_{0.3}(\%)$ & $\sigma$ & MSE & $R^2$ & Time(s) \\
\hline
LASSO  & 50.54 & 78.87 & 89.58 & 0.3480 & 0.2085  & 0.4875  & \textbf{3.5262}  \\
SVR    & 69.74 & 88.15 & 93.81 & 0.2384 & 0.1236  & 0.7442  & 1340    \\
MLP     & 78.36 & 90.85 & 95.01 & 0.2010 & 0.1032  & 0.7995  & 3953    \\
XGBoost& 77.43 & 90.81 & 95.27 & 0.2007 & 0.1015  & 0.8020  & 5628    \\
KNN    & 79.72 & 91.46 & 95.34 & 0.1924 & 0.1003  & 0.8036  & 145     \\
RF     & 79.62 & 91.39 & 95.29 & 0.1922 & 0.1010  & 0.8002  & 7970    \\
GK   &\textbf{83.65}&\textbf{92.94}&\textbf{95.56}&\textbf{0.1766}&\textbf{0.0998}&\textbf{0.8159}&146\\
\hline
\end{tabular}
\end{table*}

\begin{table*}
\caption{Performance of photometric redshift estimation of different models for the SDSS-WISE sample with 7m\_6c}
\begin{tabular}{lccccccc}
\hline
Algorithm & $\delta_{0.1}(\%)$  & $\delta_{0.2}(\%)$ & $\delta_{0.3}(\%)$ & $\sigma$ & MSE & $R^2$ & Time(s) \\
\hline
LASSO   & 50.54 & 78.87 & 89.58 & 0.3479 & 0.2085 & 0.4882 & \textbf{98}   \\
SVR     & 70.85 & 88.39 & 93.76 & 0.2365 & 0.1262 & 0.7422 & 1336 \\
MLP      & 77.11 & 90.83 & 95.24 & 0.2075 & 0.1064 & 0.7935 & 3749 \\
XGBoost & 78.83 & 91.27 & 95.44 & 0.1950 & 0.0989 & 0.8085 & 3129 \\
KNN     & 78.57 & 91.10 & 95.20 & 0.1983 & 0.1036 & 0.7956 & 282  \\
RF      & 79.76 & 91.53 & 95.37 & 0.1908 & 0.0998 & 0.8036 & 12944\\
GK &\textbf{83.25}&\textbf{92.85}&\textbf{95.61}&\textbf{0.1777}&\textbf{0.0982}&\textbf{0.8179}&319\\
\hline
\end{tabular}
\end{table*}

The comparison of predicted redshift with spectral redshift for KNN, RF, XGBoost and GeneticKNN with the input pattern 5m\_4c is shown in Figure~2. As shown in Figure~2, with 5m\_4c, the prediction has two disaster areas. Compared with the KNN, RF and XGBoost, predicted redshifts by GeneticKNN algorithm is more concentrated on the diagonal, and the scope of the disaster area is smaller.
The comparison of predicted redshift and spectral redshift for KNN, RF, XGBoost and GeneticKNN with the input pattern r\_4c is described in Figure~3. From Figure~3, with r\-4c, the prediction has two obvious disaster areas. Comparing with KNN, RF and XGBoost, GeneticKNN algorithm is more concentrated on the diagonal, and the scope of disaster area is significantly smaller. The comparison of predicted redshift and spectral redshift for KNN, RF, XGBoost and GeneticKNN with the input pattern 7m\_6c is indicated in Figure~4. GeneticKNN algorithm is more concentrated on the diagonal, and scattered points outside the boundaries $\Delta{z} = \pm{0.3}$ are rare. The comparison of predicted redshift and spectral redshift for KNN, RF, XGBoost and GeneticKNN with the input pattern r\_6c is shown in Figure~5. It is similar to that in 7m\_6c.

The distribution of the difference $\Delta{z}=z_{\rm spect}-z_{\rm reg}$ between spectral redshifts and photometric redshifts by KNN, XGBoost, RF and GeneticKNN is displayed in Figure~6. It is obvious that $\Delta{z}$ by GeneticKNN has narrower distribution than that by other algorithms with different input patterns. So GeneticKNN performs better than others. As a result, GeneticKNN is an effective and applicable method to solve the photometric redshift estimation.

\begin{figure}
\centering
\includegraphics[width=\linewidth]{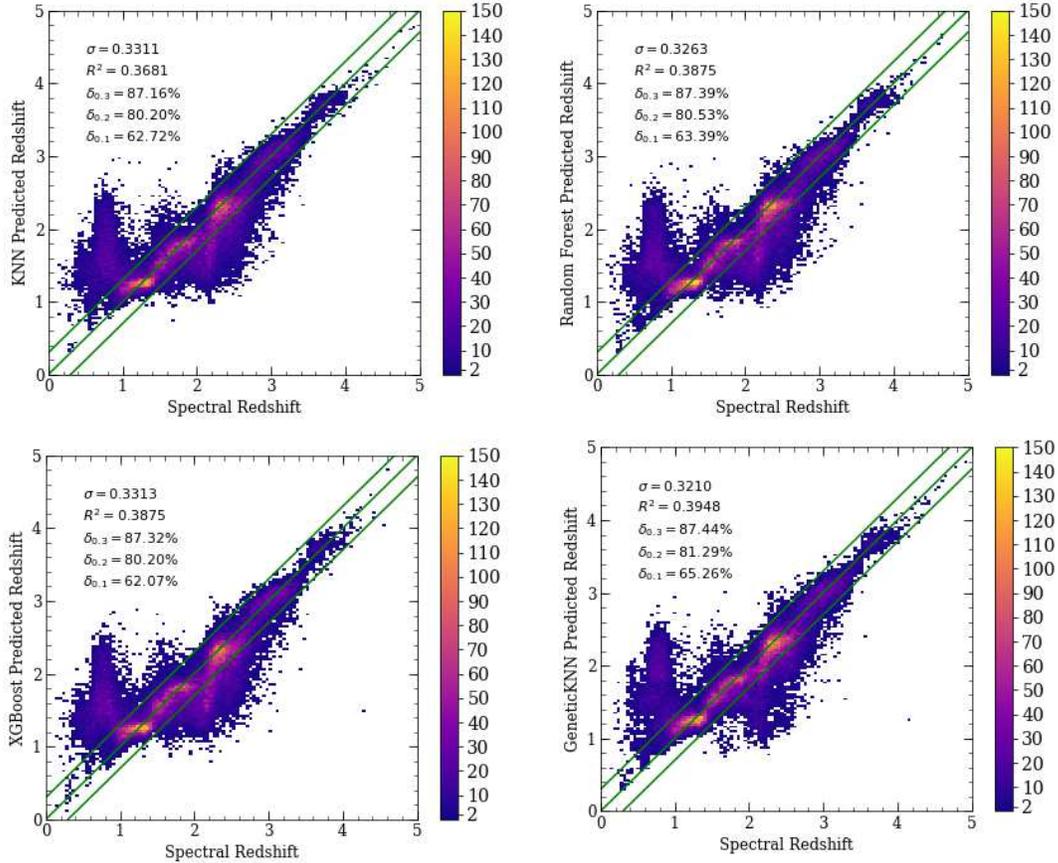}
\caption{Predicted photometric redshift vs. spectral redshift by different algorithms. The color bar shows the number of objects in each rectangle. The comparison is based on r\_4c. From left to right and from top to bottom, they are performance of KNN, RF, XGBoost and GeneticKNN, respectively. The middle line shows $z_{\rm pred}=z_{\rm reg}$ while the other two lines show $\Delta{z} = \pm{0.3}$ respectively. } \label{fig1}
\end{figure}

\begin{figure}
\centering
\includegraphics[width=\linewidth]{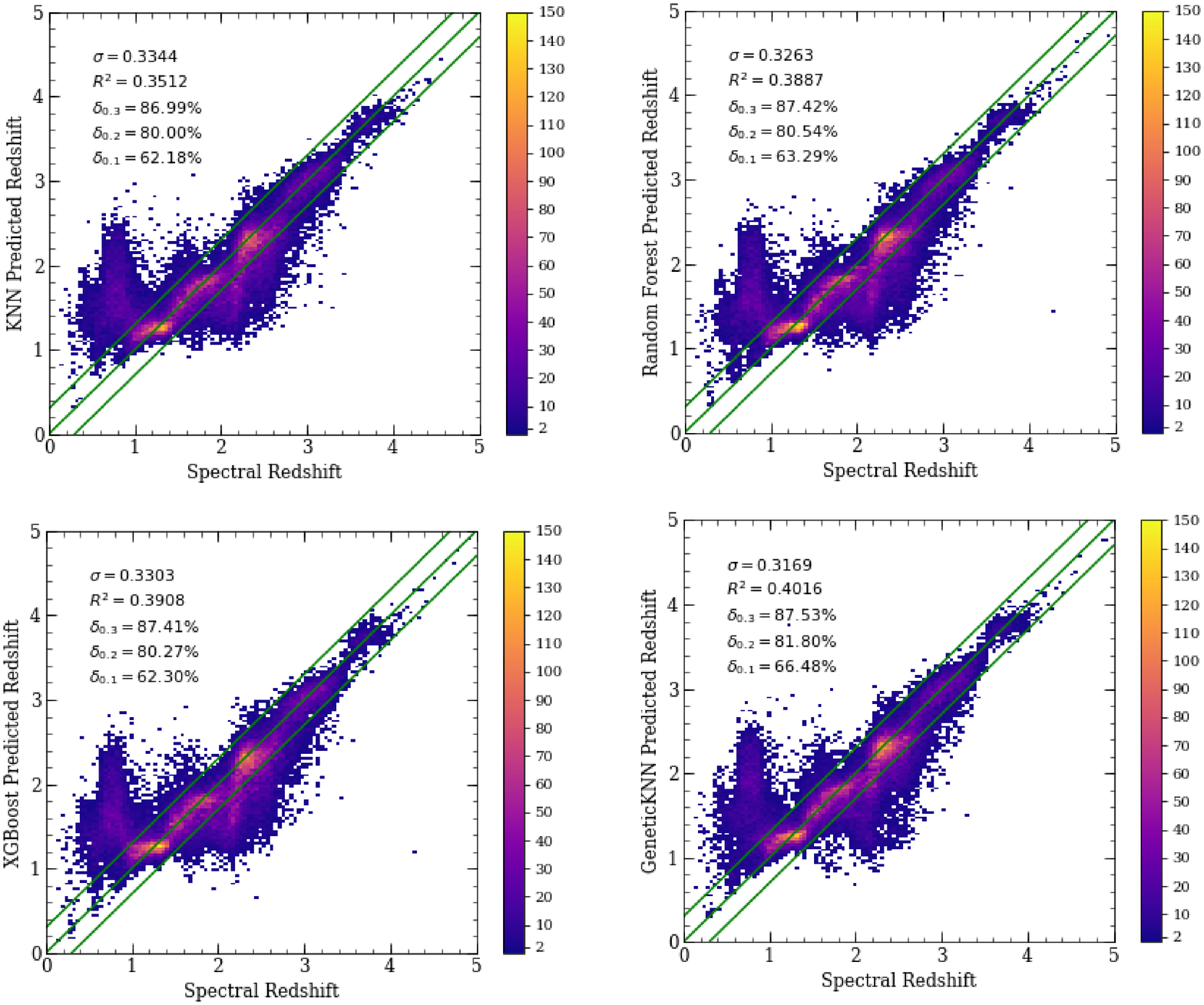}
\caption{Predicted photometric redshift vs. spectral redshift by different algorithms with 5m\_4c.} \label{fig1}
\end{figure}

\begin{figure}
\centering
\includegraphics[width=\linewidth]{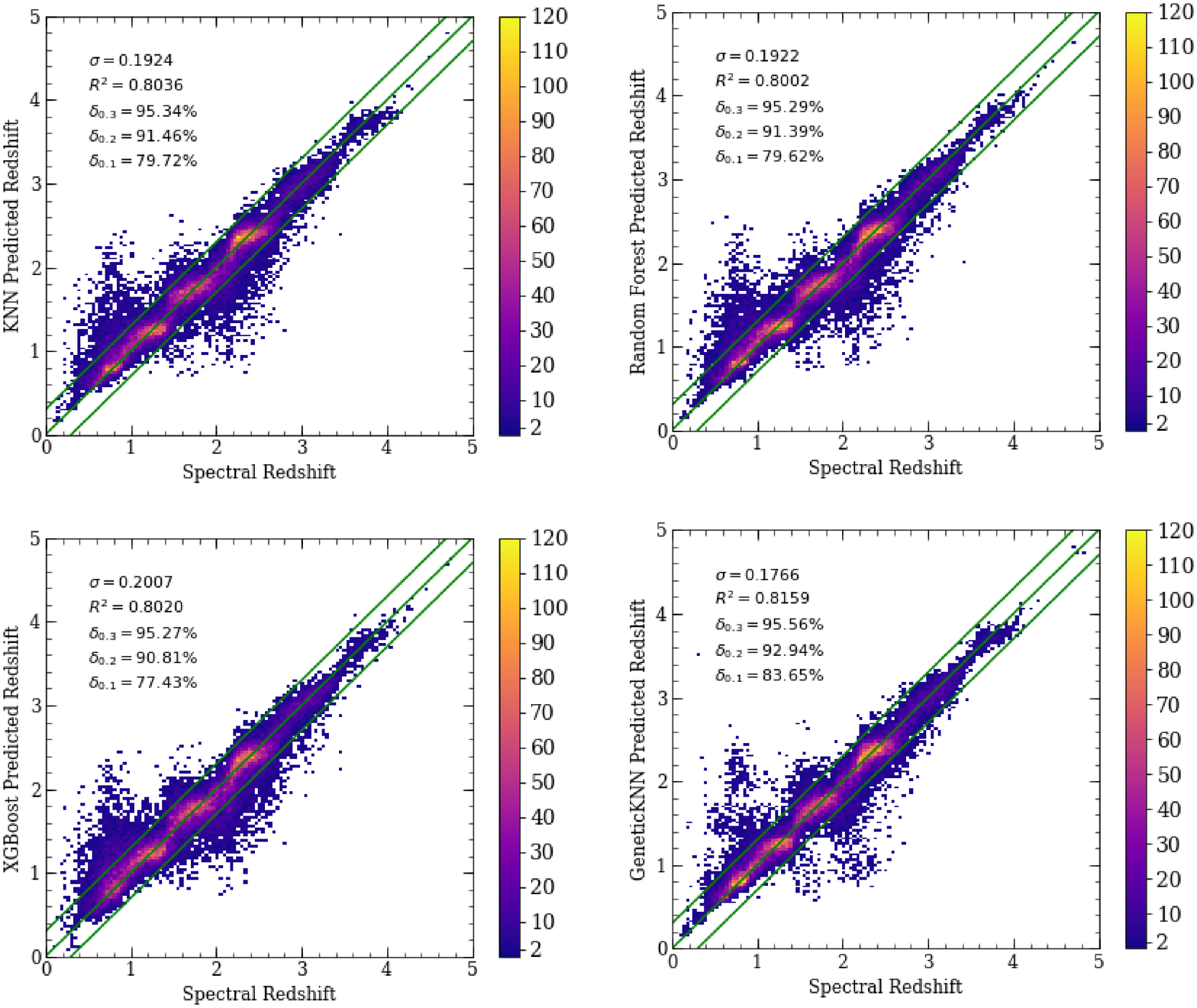}
\caption[]{Predicted photometric redshift vs. spectral redshift by different algorithms with r\_6c.} \label{fig1}
\end{figure}

\begin{figure}
\centering
\includegraphics[width=\linewidth]{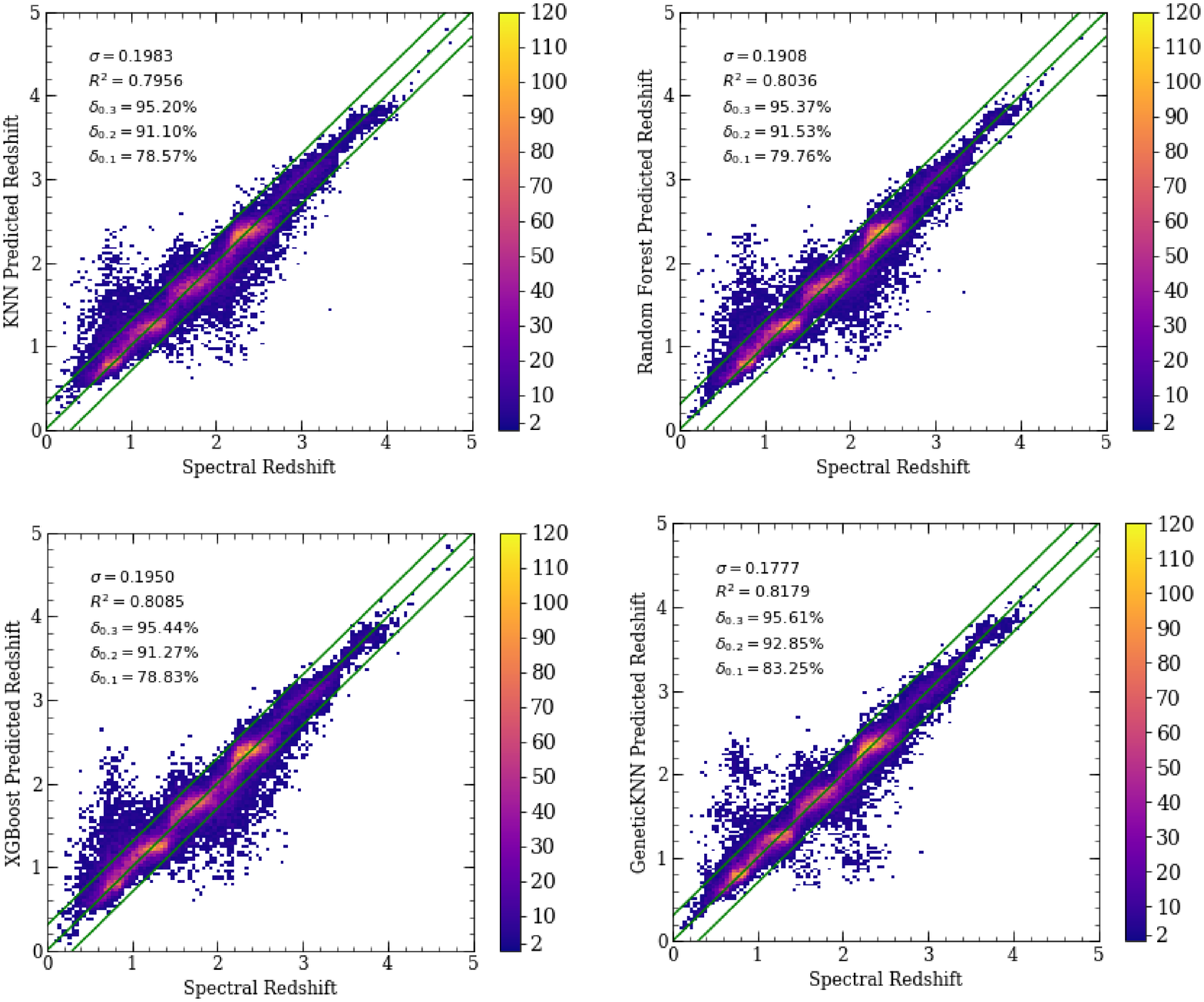}
\caption[]{Predicted photometric redshift vs. spectral redshift by different algorithms with 7m\_6c.} \label{fig1}
\end{figure}

\begin{figure}
  \centering
      \label{fig:subfig:} 
    \includegraphics[width=6cm]{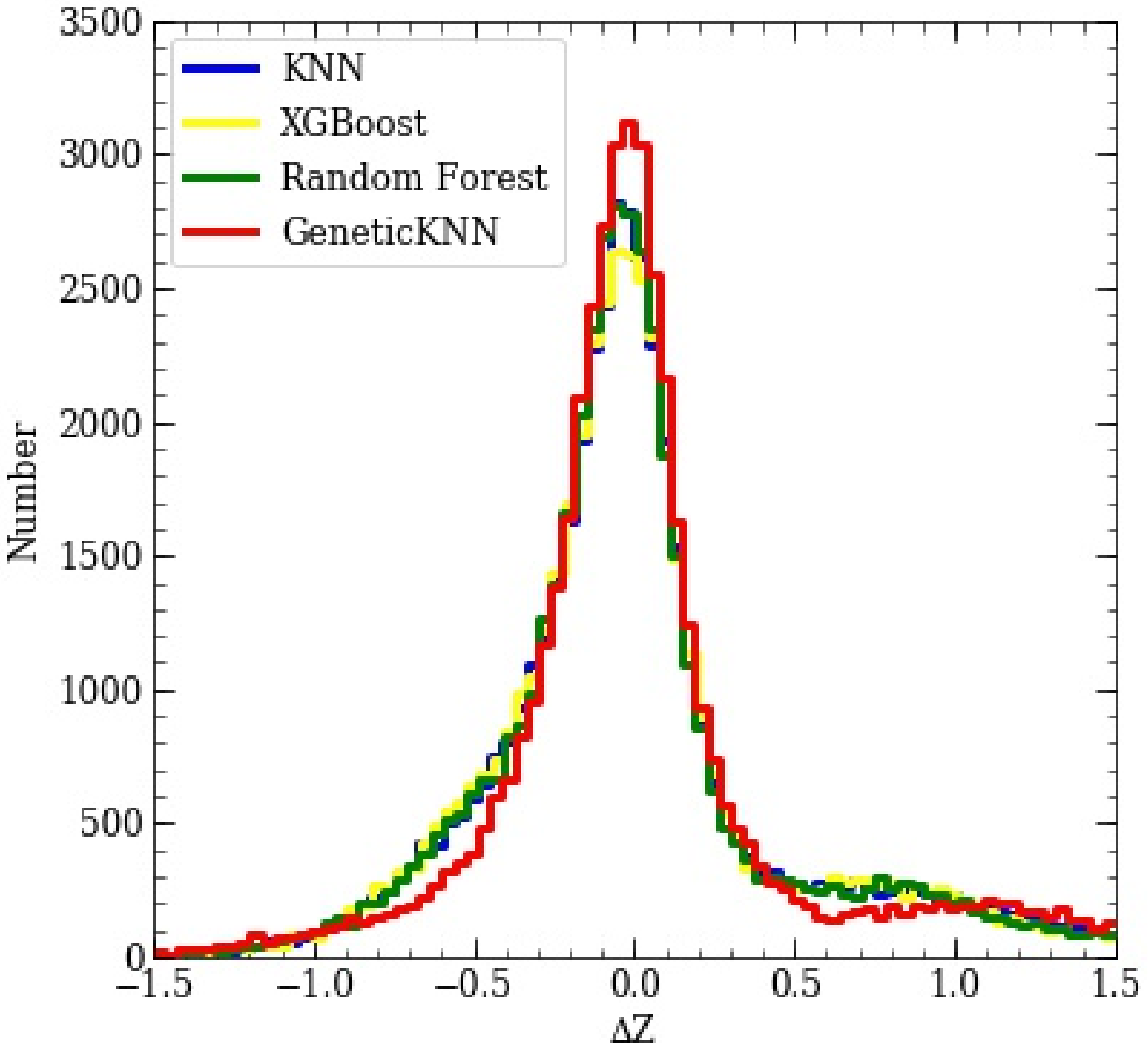}
    \includegraphics[width=6cm]{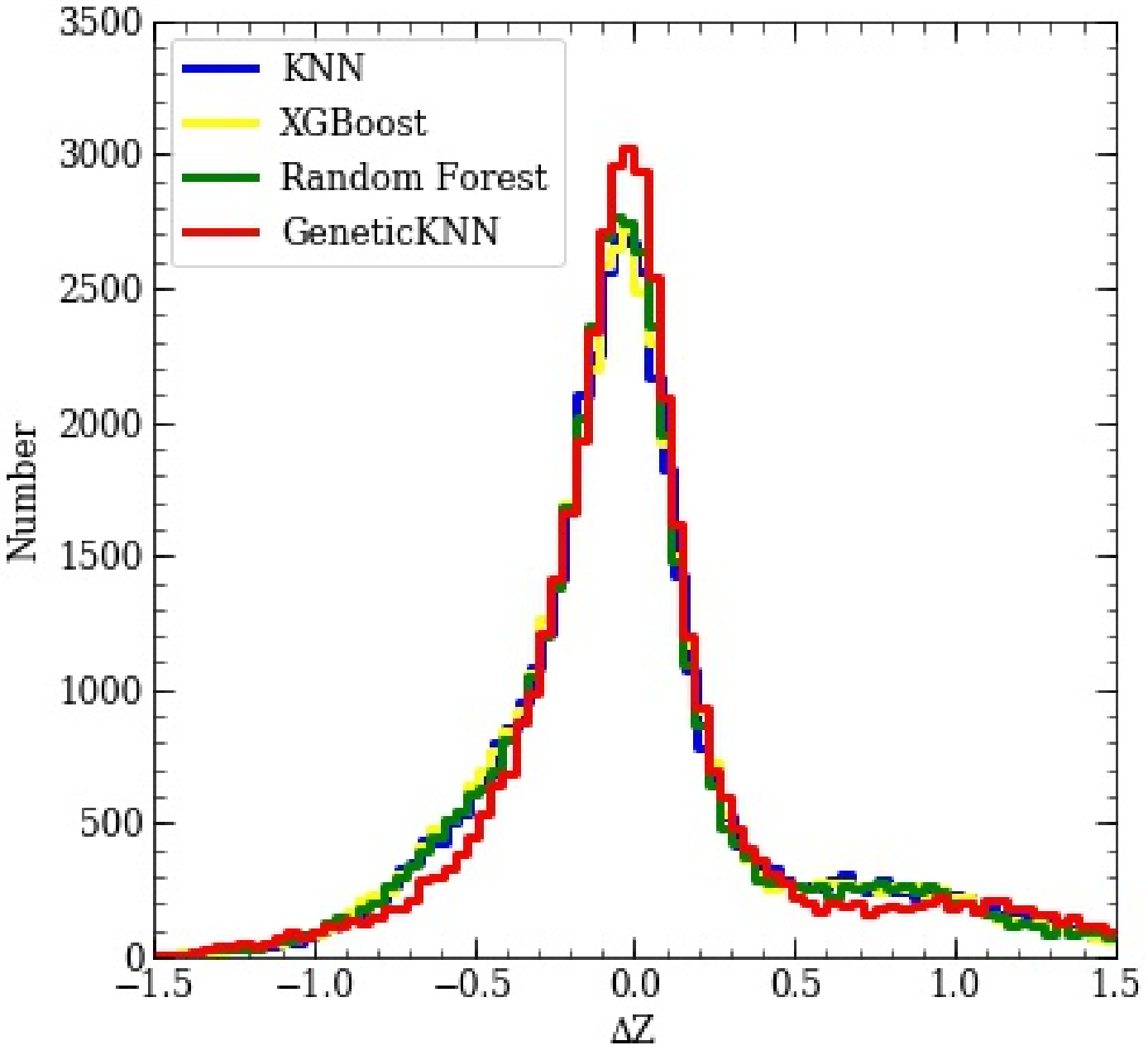}
    \includegraphics[width=6cm]{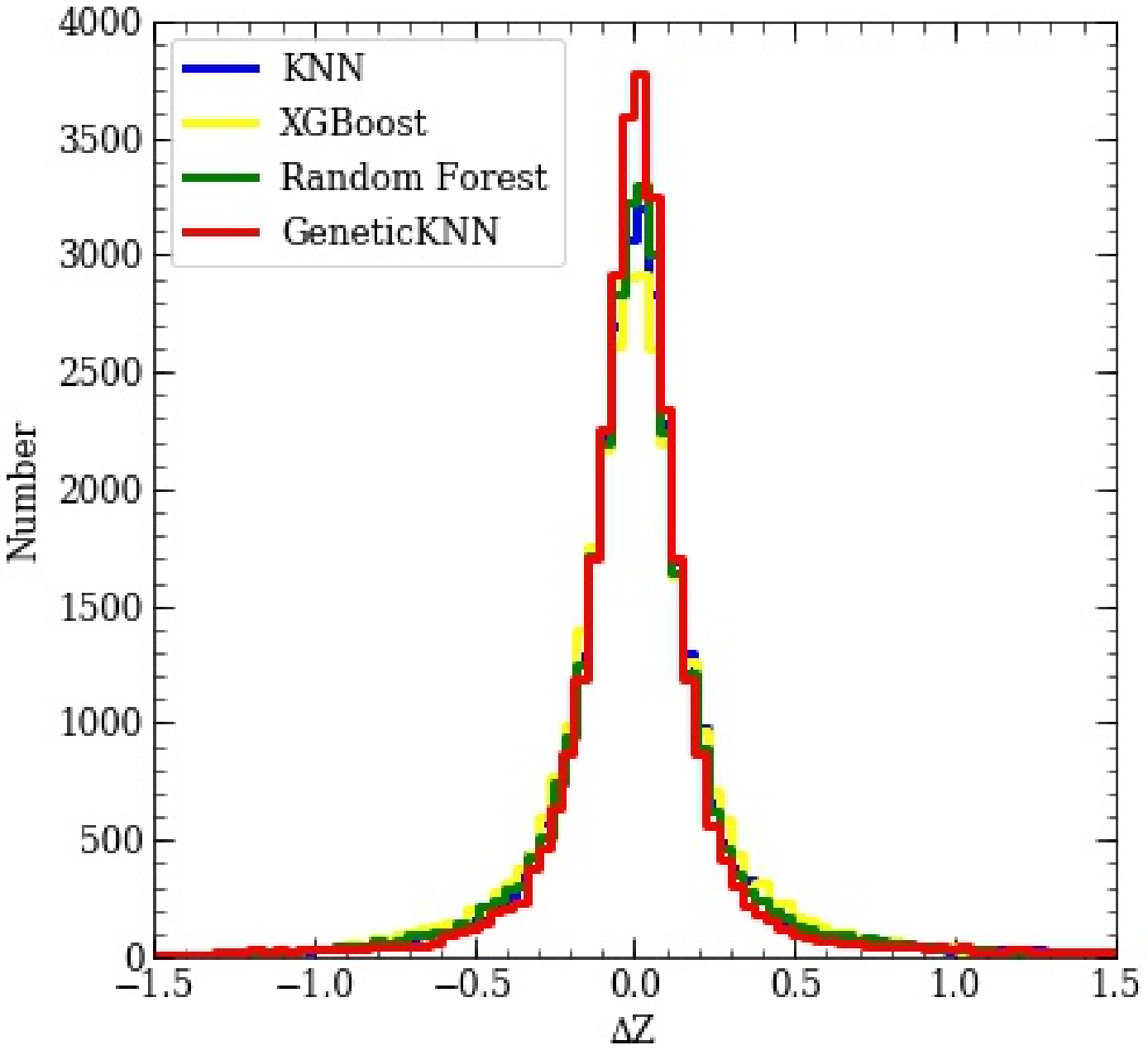}
    \includegraphics[width=6cm]{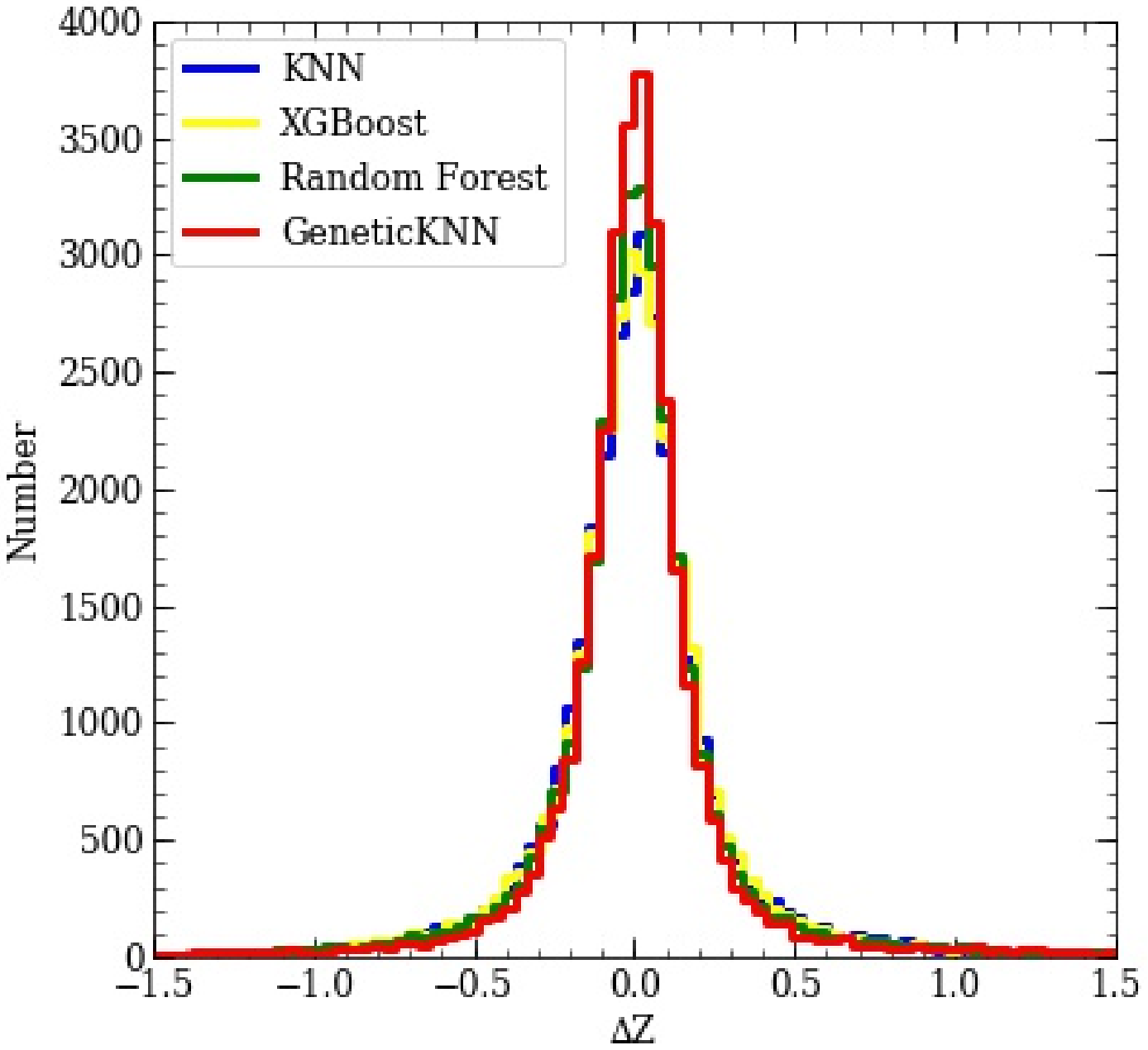}
\caption{Distribution of the difference between the spectral redshift and predicted photometric redshift for KNN, XGBoost, RF and GeneticKNN. From left to right and top to bottom, the input patterns are r\_4c, 5m\_4c, r\_6c and 7m\_6c, respectively}
\end{figure}

\section{CONCLUSION}

In the past, the focus of improving the estimation accuracy of photometric redshifts of quasars was on algorithm selection and feature selection. Perhaps the traditional normalization and dimensionality reduction methods can solve these problems, so we apply feature extraction and feature selection methods to manipulate the data, and normalize the data. For our case, the experiments show that feature extraction and feature selection methods as well as data normalization have no contribution to the performance of an algorithm for photometric redshift estimation, but reduce the estimation accuracy.

Then we compare the performance various algorithms on photometric redshift estimation. The comparison experiments indicate that KNN and random forest have a comparable performance of photometric redshift estimation. Although KNN has the advantages of simple operation, fast running time and high estimation accuracy, but it still has some disadvantages to be overcome. Therefore we put forward a new schema to improve KNN by combining genetic algorithm with KNN to optimize the data by adding weight, which is named as GeneticKNN. In general, predicted value is the average of $K$ nearest neighbors for KNN. In fact, the distance of the target from the nearest neighbors is different, the nearer points contribute more. Considering this influence, we adopt the weighted average ($p\times z_{\rm median} + (1-p)\times z_{\rm mean}$) instead of the average. The coefficient $p$ may be optimized by grid search. Based on these two improvements, the performance of GeneticKNN shows its superiority.

The experimental results show that most of evaluation indexes for GeneticKNN with the SDSS and SDSS-WISE samples have significantly increased compared to other methods. The GeneticKNN algorithm is improved in all five indexes except MSE in the SDSS sample, and all indexes in the SDSS-WISE sample are improved. Using the same method, the accuracy of the SDSS-WISE sample is better than that of the SDSS sample. As a result, our algorithm is effective and applicable. KNN algorithm is simple in principle and fast in operation, and GeneticKNN based on KNN greatly improves the estimation accuracy of photometric redshifts without increasing the running time.

In future work, we will continue to dig out more characteristics of the data itself to further improve the accuracy of photometric redshift estimation, for example, considering the piecewise correlation between features and redshift, the global correlation cannot reflect the local correlation. When applying GeneticKNN, we adjust the global correlation of features without considering the local characteristics of features. In a word, we should further improve the model from the perspective of data (e.g. feature weight by other approaches) and apply our method to other survey data (e.g. LSST).

\begin{center} \bf Acknowledgment\end{center}

We are very grateful to the refree's constructive suggestions. This paper is funded by the National Key R\&D Program of China under grant No. 2018YFB1702703, also funded by the National Natural Science Foundation of China under grant Nos.11873066, U1531122 and U1731109.

We acknowledge the use of the SDSS and WISE databases.
Funding for the Sloan Digital Sky Survey (SDSS) IV has been provided by the Alfred P. Sloan Foundation, the U.S. Department of Energy Office of Science, and the Participating Institutions. SDSS-IV acknowledges
support and resources from the Center for High-Performance Computing at
the University of Utah. The SDSS web site is www.sdss.org.
SDSS-IV is managed by the Astrophysical Research Consortium for the
Participating Institutions of the SDSS Collaboration including the
Brazilian Participation Group, the Carnegie Institution for Science,
Carnegie Mellon University, the Chilean Participation Group, the French Participation Group, Harvard-Smithsonian Center for Astrophysics,
Instituto de Astrof\'isica de Canarias, The Johns Hopkins University,
Kavli Institute for the Physics and Mathematics of the Universe (IPMU) /
University of Tokyo, the Korean Participation Group, Lawrence Berkeley National Laboratory,
Leibniz Institut f\"ur Astrophysik Potsdam (AIP),
Max-Planck-Institut f\"ur Astronomie (MPIA Heidelberg),
Max-Planck-Institut f\"ur Astrophysik (MPA Garching),
Max-Planck-Institut f\"ur Extraterrestrische Physik (MPE),
National Astronomical Observatories of China, New Mexico State University,
New York University, University of Notre Dame,
Observat\'ario Nacional / MCTI, The Ohio State University,
Pennsylvania State University, Shanghai Astronomical Observatory,
United Kingdom Participation Group,
Universidad Nacional Aut\'onoma de M\'exico, University of Arizona,
University of Colorado Boulder, University of Oxford, University of Portsmouth,
University of Utah, University of Virginia, University of Washington, University of Wisconsin,
Vanderbilt University, and Yale University. The Wide-field Infrared Survey Explorer (WISE) is
a joint project of the University of California, Los Angeles, and the Jet Propulsion Laboratory/California
Institute of Technology, funded by the National Aeronautics and Space Administration.

\begin{center}\bf{Appendix}\end{center}

\begin{appendix}

\section{Consideration and Optimization of Genetic Strategy for Feature Weight Search}

The genetic algorithm firstly needs to analyze the problem deeply, determines the solution space of the problem, and encodes the solution space, in which each code corresponds to a solution space. Then it selects a fixed number of solutions randomly as the initial solution set, and calculates the adaptive values of all solutions in the solution set according to the selected fitness function. Secondly, according to the survival of the fittest rule, some solutions are selected as candidate sets, and crossover and mutation of candidate sets are performed according to the crossover rate and mutation rate specified in advance to generate a new generation of solution sets. Finally, it uses the fitness function to calculate the corresponding fitness value for all the solutions in the new generation solution set again, and the loop iterated through these steps until the iteration termination condition is satisfied. The approximate optimal solution of the problem can be obtained by decoding the best solution of the obtained solution set.

\begin{enumerate}
\item DESIGN OF INDIVIDUAL INITIALIZATION OF FEATURE WEIGHT COMBINATION

The primary operation of genetic algorithm is to initialize the population individuals. Individuals are the basic structure of a population. A large number of individuals constitute a complete population. In the genetic algorithm, individuals can also be called gene sequences or chromosomes, and individual design is also called individual coding. In the problem of weight search, the ultimate goal of genetic algorithm is to find the optimal corresponding weight of all features, so in this problem, each individual is the corresponding weight list, such as [w$_{1}$, w$_{2}$, $\cdots$, w$_{n}$].
For the 5m\_4c input pattern of the SDSS dataset, it has nine features, so the corresponding weight list is [w$_{1}$, w$_{2}$, $\cdots$, w$_{9}$]; for r\_4c, it has five features, so the corresponding weight list is [w$_{1}$, w$_{2}$, $\cdots$, w$_{5}$]. And in the SDSS-WISE dataset, for 7m\_6c input pattern, there are thirteen features and the corresponding weight list is [w$_{1}$, w$_{2}$, $\cdots$, w$_{13}$]; for r\_6c, the corresponding weight list is [w$_{1}$, w$_{2}$, $\cdots$, w$_{7}$]. Considering that the influence of features on the model is not always positive, the corresponding values of weight is between the range of [-10, 10].

\item COMPARISON AND SELECTION OF GENETIC FITNESS BASED ON KNN

In genetic algorithm, each individual calculates its own fitness. According to the arrangement of fitness from large to small, individuals with higher fitness have the right to be selected first. Therefore, the selection of fitness will determine the speed of optimization of genetic algorithm. In the meantime, the selection of fitness also represents the main direction of model optimization. In this experiment, the following indicators are used to represent fitness.

\begin{enumerate}

\item $\delta_{0.3}$: Redshift normalized residual (when $e=0.3$), that is, the rate of the absolute error between the predicted value and the true value which is less than $0.3\times $(1+true value). The aim of choosing $\delta_{0.3}$ as the fitness is to let more predictions fall into the acceptable range. But the problem is that it only considers the majority of the samples but ignores whether the predicted values deviate from the real values, which will lead to the excessively deviated samples.

\item Mean absolute error (MAE): The mean of the absolute error between the truth value and the estimated value of all samples. Choosing MAE as the fitness could avoid offsetting errors and reduce the absolute error of the model concisely and effectively.

\item Mean square error (MSE): The square root mean of the absolute error between the truth value and the estimated value. Compared with MAE, MSE is better at capturing the effects of outliers.
\end{enumerate}

Different settings have certain influence on the weight generation. The selection of fitness can be based on the requirements of the actual dataset. In our study, the SDSS quasar sample is prone to prediction disasters, so we select MSE as the fitness considering that excessive deviation of abnormal data should be reduced appropriately; for the SDSS-WISE quasar sample, we mainly consider the overall absolute error, so we selected MAE as the fitness function.

\item BUILD SELECTION LAYER BASED ON ELITE STRATEGY AND TOURNAMENT

In the principle of genetic algorithm, the selection strategy is particularly important, which determines the individuals participating in the crossover operation in the current generation and directly affects whether the genes of the current individual can be inherited to the next generation. In our study, we use Elite Strategy and Tournament Selection Method to construct the selection layer of genetic algorithm.

\begin{enumerate}

\item Elite Strategy. The purpose of genetic algorithm is to generate the optimal fitness individuals, and in the face of selection phase, it is easy to select and cross the current batch of better individual genes, which is easy to lead to the loss of excellent individual gene combinations. Elite strategy is an optimization strategy for basic genetic algorithm. To prevent the selection of the best solution and the disruption of the mutation process, we directly copy a batch of superior individuals from each generation to the next generation. At the same time, since the size of the population remains unchanged, a certain amount of elites are directly added to the new generation group, which means that in the process of producing the next generation of the population, some individuals with low adaptability will be eliminated by the new generation population due to poor adaptability. A large number of elite individuals may be trapped in a local optimal solution. The elite in the current generation may not be the grandfather of the final optimal solution, the direct reservation of the elite will lead to the exclusion of other potential possibilities, and the similarity of the elite will also lead to the lower value of the crossover process. In this weight search, in order to avoid the algorithm falling into the local optimal solution, the number of elites would be limited to 10\% of the population.

\item Tournament Selection. As the name suggests, Tournament Selection is a tournament in which two or more individuals are selected randomly. The highly adaptable individuals are selected as the winner, and this process is repeated until the entire selection process is completed. The method is more random but has the process of comparing and winning. This process ensures that the excellent individuals will be retained while the poor ones will be eliminated (excellent individuals can be guaranteed to be retained as long as they are selected into the tournament, while poor individuals will be eliminated even if they are selected into the tournament). In order to avoid the situation that excellent individuals are not selected into the tournament, the size of a single tournament is set at 4 in this experiment. That means, 4 individuals would be randomly selected to compete in each tournament, so as to improve the chance of excellent individuals being selected.
\end{enumerate}

\item REASONABLE VARIATION STRATEGY OF WEIGHT COMBINATION

The selection of mutation strategy in genetic algorithm will determine the change of the current population. If there is no optimal solution in the current population, the mutation strategy will directly determine whether genetic algorithm can finally find the optimal solution. In the process of mutation, two mutation strategies will be considered.

\begin{enumerate}

\item Expand the search for peripheral values. The current individual may not be the optimal value of the surrounding, and the appropriate deviation in the surrounding may produce a better individual. Based on this fact, the variation strategy considers that the current data changes by 0.9 times or 1.1 times, so as to make the individual variation produce some small changes.

\item Random valid value substitution. The current individual may be trapped in the dilemma of local optimum, and appropriate reference to other individuals may help the current individual to escape from this local area. Inspired by this, the variation strategy would consider replacing part of the values of some individuals with the corresponding parameter values of other random individuals, and the effective values of other individuals may also be applicable to the current individuals. Under this strategy, the individual variation of relatively large changes not only has been achieved, but the unreasonable variation can also be avoided to a certain extent.
\end{enumerate}
\end{enumerate}

\section{Estimation Optimization Based on Nearest Neighbor Value}

After obtaining weights from genetic algorithms, we apply $K$-Nearest Neighbor (KNN) method to search for most similar records for estimation based on weighted attribute distance computation. Usually, KNN method selects the arithmetic mean of the estimated values of $K$ nearest neighbors to predict the target. However, the mean value will introduce a large error into the estimation if there are some extreme values in the $K$ neighbors. Meanwhile, the median isn't affected by the extremes, but if the median replaces the arithmetic mean, the distribution of the estimated values is not considered. Thereby, we sort the real values of the $K$ nearest neighbors and check whether the value distribution is sparse and dense. Next, we apply a weighted average of the mean and the median value of the nearest neighbors as the estimation. The weight here is dependent on neighbor distribution. By observation, the estimation accuracy of target value is generally higher in the region with dense estimated value distribution, and lower in the part with sparse estimated value distribution. Therefore, the distribution density and location of the nearest neighbor estimation should be analyzed. In this way, the weight can be selected reasonably.

Assuming that the coefficient is $p$, the predicted redshift is $p\times z_{\rm median} + (1-p)\times z_{\rm mean}$, where $z_{\rm median}$ is the redshift median of the $K$ neighbors and $z_{\rm mean}$ is their redshift average of the $K$ neighbors.

\section{Evaluation Index for Photometric Redshift Estimation}

In order to compare the fitting degree of different models, we can analyze different indicators, such as  MSE, MAE, R$^{2}$ and so on. We can compare models by comparing them against each other and against all possible sub-models. Photometric redshift estimation is continuous variable prediction, which belongs to the regression task, so the algorithm suitable for regression is able to be applied to estimate photometric redshifts. Given the sample, it is needed to select the regression method. Comparison of different regressors always depends on different regression indexes. Another criterion for determining the goodness of photometric redshift estimation is to satisfy $|\Delta{z}| = |z_i -\hat{z}_i| < e$.

The definition of mean absolute error (MAE) $\sigma$ is:

\begin{equation}\label{1}
\sigma= \frac {1}{n}\sum ^{n}_{i=0}\left| z_{i}-\hat{z}_{i}\right|
\end{equation}
where $z_i$ is the true value of redshift, $\hat{z}_i$ is the estimated value of redshift, $n$ is the sample size.
The fraction of samples that satisfy  $|\Delta{z}| = |z_i -\hat{z}_i| < e$ is usually used to evaluate the effect of photometric redshift estimation, where $e$ is the given residual threshold (Schindler et~al. 2017).
\begin{equation}\label{2}
f_{|\Delta{z}|<e} = \frac{N(|z_i-\hat{z}_i|<e)}{N_{\rm total}}
\end{equation}
The typical values of $e$ is 0.1, 0.2 and 0.3. But in fact, we usually adopt the redshift normalized residuals,
\begin{equation}\label{3}
\delta_{e}= \frac{N(|z_i-\hat{z}_i|< e(1+ z_i))}{N_{\rm total}}
\end{equation}

The mean square error (MSE) is the expectation of the square of the difference between the estimated value and the true value. It can be used to evaluate average error and the variation degree of data. The smaller the MSE is, the better the predictability of the predictive model can be described.
\begin{equation}\label{4}
MSE = \frac{1}{n} \sum ^{n}_{i=0} (z_i - \hat{z}_i)^2
\end{equation}
The total sum of squares (TSS) is:
\begin{equation}\label{5}
TSS = \sum ^{n}_{i=0} (z_i - \overline{z}_i)^2
\end{equation}
The regression sum of squares (ESS) is:
\begin{equation}\label{6}
ESS = \sum ^{n}_{i=0} (\hat{z}_i - \overline{z}_i)^2
\end{equation}
The residual sum of squares (RSS) is:
\begin{equation}\label{7}
RSS = \sum ^{n}_{i=0} (z_i - \hat{z}_i)^2
\end{equation}
Their relation is $TSS = ESS + RSS$.
The observed values of $y$ can be divided into two parts: one part comes from ESS and the other part comes from RSS. When TSS remains unchanged, with a given sample, the closer the true value is to the regression line of the sample, the larger the proportion of ESS in TSS is. Therefore, the fitting degree is defined as the ratio of ESS to TSS.
\begin{equation}\label{8}
R^2 = \frac{ESS}{TSS} = 1 - \frac{RSS}{TSS}
\end{equation}
We also may use $R^2$ to depict the performance of regression.
\begin{equation}\label{9}
R^2 = 1- \frac{\sum ^{n}_{i=0} (z_i - \hat{z}_i)^2}{\sum ^{n}_{i=0} (z_i - \overline{z}_i)^2}
\end{equation}

For the obtained sample data, $\sum^{n}_{i=0} (z_i - \hat{z}_i)^2$ of $R^2$ is fixed. Therefore, the larger $R^2$ is, the smaller the RSS is, that means the better the fitting effect of the model is. In the linear regression model, $R^2$ represents the contribution rate of explanatory variables to the change of prediction variables. The closer $R^2$ is to 1, the better the regression effect is. Time is the total time of training and prediction process by 10-fold cross validation. For 10-fold cross validation, we separate the training sample into 10 subgroups, each experiment is done with 9 subgroups for training and the left one for test in turn for ten times, keeping any subgroup for test.

\end{appendix}

\end{document}